\documentclass[a4paper, 12pt]{article}

\usepackage{amsmath}
\usepackage{graphics}

\begin{document}
\renewcommand{\theequation}{\thesection .\arabic{equation}}

\newcommand{\sign}{\operatorname{sign}}
\newcommand{\Ci}{\operatorname{Ci}}
\newcommand{\tr}{\operatorname{tr}}

\newcommand{\beq}{\begin{equation}}
\newcommand{\eeq}{\end{equation}}
\newcommand{\beqn}{\begin{eqnarray}}
\newcommand{\eeqn}{\end{eqnarray}}

\newcommand{\slp}{\raise.15ex\hbox{$/$}\kern-.57em\hbox{$ \partial $}}
\newcommand{\lnA}{\raise.15ex\hbox{$/$}\kern-.57em\hbox{$A$}}
\newcommand{\unmedio}{{\scriptstyle\frac{1}{2}}}
\newcommand{\uncuarto}{{\scriptstyle\frac{1}{4}}}

\newcommand{\trial}{_{\text{trial}}}
\newcommand{\true}{_{\text{true}}}
\newcommand{\const}{\text{const}}

\newcommand{\intp}{\int\frac{d^2p}{(2\pi)^2}\,}
\newcommand{\intx}{\int d^2x\,}
\newcommand{\inty}{\int d^2y\,}
\newcommand{\intxy}{\int d^2x\,d^2y\,}

\newcommand{\bP}{\bar{\Psi}}
\newcommand{\bc}{\bar{\chi}}
\newcommand{\hs}{\hspace*{0.6cm}}

\newcommand{\bra}{\left\langle}
\newcommand{\ket}{\right\rangle}
\newcommand{\bracket}{\left\langle\,\right\rangle}

\newcommand{\D}{\mbox{$\mathcal{D}$}}
\newcommand{\N}{\mbox{$\mathcal{N}$}}
\newcommand{\Lag}{\mbox{$\mathcal{L}$}}
\newcommand{\V}{\mbox{$\mathcal{V}$}}
\newcommand{\Z}{\mbox{$\mathcal{Z}$}}

\begin{titlepage}


\vspace{2cm}

\begin{center}

{\Large {\bf RG study of a non local sine-Gordon model}}

\vspace{1.3cm}

Carlos M. Na\'on$^{a,b}$ and Mariano J.
Salvay$^{a,b}$\footnote{e-mail: salvay
@fisica.unlp.edu.ar,
naon@fisica.unlp.edu.ar}

\vspace{.8cm}

$^a$ {\it Instituto de F\'{\i}sica La Plata, Departamento de
F\'{\i}sica, Facultad de Ciencias Exactas, Universidad Nacional de
La Plata.  CC 67, 1900 La Plata, Argentina.}

\smallskip

$^b$ {\it Consejo Nacional de Investigaciones Cient\'{\i}ficas y
T\'ecnicas, Argentina.}

\vspace{1cm}

\begin{abstract}
We study a non local version of the sine-Gordon
model connected to a many-body system with backward and umklapp
scattering processes. Using renormalization group methods we
derive the flow equations for the couplings and show how non
locality affects the gap in the spectrum of charge-density
excitations. We compare our results with previous predictions
obtained through the self-consistent harmonic approximation.
\end{abstract}
\end{center}

\vspace{1 cm}


\noindent{\it Pacs:} 11.10.Lm, 05.30.Fk

\end{titlepage}

\newpage
\section{Introduction}
\hs In general, Quantum Field Theories have been built in the
context of local models. However, there exist physical situations
that lead to non local interactions in a straightforward way. Let
us mention, for instance, Wheeler and Feynman's description of
charged particles \cite{Wheeler}, string theories with non local
vertices \cite{strings}, and non local kinetic terms that appear
when bosonizing fermions in (2+1) dimensions \cite{Marino}
\cite{Barci}. As we shall see, several recently considered non
local field theories are related to the study of electronic
systems in one spatial dimension (1D) \cite{Krive}
\cite{Nosotros1}. Indeed, in recent years the physics of 1D
systems of strongly correlated particles has become a very
interesting subject since one can take advantage of the simplicity
of the models at hand and, at the same time, expect to make
contact with experiments. For instance, the recently discovered
carbon nanotubes are perfect experimental realizations of 1D
conductors \cite{nano}. On the other hand, as the dimensionality
of a system decreases, charge screening effects become less
important and the long-range interaction between electrons is
expected to play a central role in determining the properties of
the system. In fact, from a theoretical point of view the effects
of long-range interactions have been recently discussed in
connection to several problems such as the Fermi-edge singularity
\cite{Fermi-edge}, the insulator-metal transition
\cite{insulator-metal}, the role of the lattice through umklapp
scattering and size dependent effects \cite{lattice}, etc. In the
specific context of carbon nanotubes, a low-energy theory
including Coulomb interactions has been also recently derived and
analyzed \cite{EG}. Non local fermionic models have been also used
in the study of fluctuation effects in low-dimensional
Spin-Peierls systems \cite{SP}.

As shown in \cite{Nosotros2}, starting from a non-local and
non-covariant version of the Thirring model \cite{Thirring}, in
which the fermionic densities and currents are coupled through
bilocal, distance-dependent potentials, one can make direct
contact with the "g-ology" model \cite{g-ology} currently used to
describe different scattering processes characterized by coupling
functions $g_1$, $g_2$, $g_3$ and $g_4$. When one bosonizes this
theory by either operational or functional methods, due to the
contributions of $g_1$ (backscattering) \cite{back} and $g_3$
(umklapp) \cite{um} one finds an even more drastic departure from
the local case. Indeed, instead of the well-known integrable
sine-Gordon model (SG) one gets a non local extension of it,
which, as far as we know, is not exactly solvable. Recently, in
ref.\cite{Nosotros2}, the physical content of this model was
explored by using the self-consistent harmonic approximation
(SCHA) \cite{scha}. As it is well-known the SCHA is a non
controlled approximation, i.e. there is no perturbative parameter
involved. It is then desirable to have an alternative analysis of
this problem. This is the main motivation of the present work. We
will apply the renormalization group (RG) technique
\cite{Shankar}, usually employed in local cases, to the non local
sine-Gordon model (NLSG) mentioned above. For simplicity we shall
assume that non locality plays a role only in umklapp interactions
($g_3$) whereas all other potentials are local, i.e. proportional
to delta functions. In Section 2 we briefly show how the NLSG
action is obtained from the non local Thirring model. In Section 3
we derive the RG equations and compute the gap in the
charge-density spectrum. This allows us to determine the effect of
non contact $g_3$ couplings. Finally, in Section 4, we discuss our
results.

\section{The Model}
\setcounter{equation}{0}

\hs Let us sketch the derivation of the NLSG action. We start from
the fermionic (1+1)-dimensional Quantum Field Theory with
Euclidean action given by

\begin{equation}
S = S_0 + S_{fs} + S_{bs} + S_{us}
\label{1}
\end{equation}
where

\begin{equation}\label{2}
S_0 = \intx\bP i\slp\Psi
\end{equation}
is the unperturbed action associated to a linearized free
dispersion relation. The contributions of the different scattering
processes can be written as

\begin{equation}
S_{fs} = - \frac{g^2}{2} \intxy ( \bP \gamma_{\mu} \Psi ) (x)~
V_{(\mu)}(x,y) ~( \bP \gamma_{\mu} \Psi ) (y) \label{3}
\end{equation}
and

\begin{equation}
S_{bs} + S_{us}= -\frac{{g'}^2}{2} \intxy (\bP~\Gamma_{\mu}
\Psi)(x)\,U_{(\mu)}(x,y)(\bP~\Gamma_{\mu} \Psi)(y) \label{4}
\end{equation}
where the $\gamma_{\mu}'s$ are the usual two-dimensional Dirac
matrices and $\Gamma_{0}=1$, $\Gamma_{1}=\gamma_{5}$. The coupling
potentials $V_{(\mu)}$ and $U_{(\mu)}$ are assumed to depend on
the distance $\mid x-y \mid$ and can be expressed in terms of
"g-ology" parameters as

\begin{align}
V_{(0)}(x,y) &= \frac{1}{g^2}(g_2+g_4)(x,y) \\ V_{(1)}(x,y) &=
\frac{1}{g^2}(g_2-g_4)(x,y) \\ U_{(0)}(x,y) &= \frac{1}{{g'}^2}(g_3+g_1)(x,y) \\
U_{(1)}(x,y) &= \frac{1}{{g'}^2}(g_3-g_1)(x,y)
\end{align}
In the above equations $g$ and ${g'}$ are just numerical constants that could be set
equal to one. We keep them to facilitate comparison of our results with those
corresponding to the usual Thirring model. Indeed, this case is obtained by choosing
${g'}=0$ and $V_{(0)}(x,y)=V_{(1)}(x,y)=\delta^2(x-y)$. On the other hand, the
non-covariant limit ${g'}=0$, $V_{(1)}(x,y)=0$ gives one version ($(g_2 =g_4$) of the
TL model \cite{TL}.

The terms in the action containing $g_2$ and $g_4$ represent forward scattering
events, in which the associated momentum transfer is small. In the $g_2$ processes the
two branches (left and right-moving particles) are coupled, whereas in the $g_4$
processes all four participating electrons belong to the same branch. On the other
hand, $g_1$ and $g_3$ are related to scattering diagrams with larger momentum
transfers of the order of $2 k_F$ (bs) and $4 k_F$ (us) respectively (this last
contribution is important only if the band is half-filled). For simplicity, throughout
this paper we will consider spinless electrons. The extension of our results to the
spin-$1/2$ case with spin-flipping interactions, though not trivial, could be done by
following the lines of ref. \cite{spinflipping}.

At this point we consider the partition function $\Z$ expressed as
a functional integral over fermionic variables. The implementation
of a generalized Hubbard-Stratonovich identity \cite{Nosotros1}
allows to write $\Z$ in terms of a fermionic determinant. Although
this determinant is highly non trivial, one can combine a chiral
change in the fermionic path-integral measure with a formal
expansion in $g'$ in order to obtain a bosonic representation (See
ref. \cite{Nosotros2} for details). One thus establishes an
equivalence between the original fermionic action and the
following bosonic action depending on five scalars $\Phi$, $\eta$,
$C_0$, $C_1$ and $\varphi$:
\begin{multline}
S_{bos}=\intp [ \Phi (p) \Phi (-p) A(p) + \eta (p) \eta (-p) B(p)
+ \Phi (p) \eta (-p)C(p) +\varphi (p)\varphi (-p)\frac{p^2}{2}]\\
+\frac{1}{2}\intxy C_{\mu} (x) U^{-1}_{(\mu)}(x,y) C_{\mu} (y)
+\frac{{g'} \Lambda c}{\pi} \intx  [ C_{(0)} (x)f_{0}(x) + i
C_{(1)} (x)f_{1}(x)] \label{39}
\end{multline}
where
\begin{align}
A(p)&= \frac{1}{2}\left[p_0 ^2 \hat{V}_{(1)}^{-1} (p) + p_1 ^2 (\hat{V}_{(0)}^{-1} (p)
+ \frac{g^2}{\pi})\right] \\ B(p)&=\frac{1}{2}\left[p_0 ^2 \hat{V}_{(0)}^{-1} (p) +
p_1 ^2 (\hat{V}_{(1)}^{-1} (p) - \frac{g^2}{\pi})\right] \\
C(p)&=p_0p_1\left(\hat{V}_{(0)}^{-1} (p)- \hat{V}_{(1)}^{-1} (p) +
\frac{g^2}{\pi}\right),
\end{align}
and
\begin{align}
f_{0}(x)=&\cos ((\sqrt{4 \pi}\varphi -2i g \Phi )(x))\\
f_{1}(x)=& \sin (( \sqrt{4 \pi}\varphi - 2i g \Phi )(x))
\label{40}
\end{align}

Since the integrals in $C_0$ and $C_1$ are quadratic these fields are easily
integrated out and one gets

\begin{equation}
\Z = \N \int \mbox{$\mathcal{D}$} \Phi ~ \mbox{$\mathcal{D}$} \eta
~ \mbox{$\mathcal{D}$} \varphi ~ e^{-S_{eff}[\Phi,\eta,\varphi]}
\label{41}
\end{equation}
with

\begin{equation}
S_{eff}[\Phi,\eta,\varphi ] = S_{0} + S_{int}
\label{42}
\end{equation}
where

\begin{equation}
S_{0}=\intp [ \Phi (p) \Phi (-p) A(p) + \eta (p) \eta (-p) B(p)
 +  \Phi (p) \eta (-p) C(p) + \varphi (p)\varphi
(-p)\frac{p^2}{2}]
\end{equation}

\begin{multline}\label{43}
S_{int}=-\frac{(\Lambda c)^2}{2\pi^2}\intxy g_{1} (x,y) \cos\left[
\sqrt{4\pi}(\varphi(x) -\varphi (y)) -2i g (\Phi(x) -\Phi(y))\right]\\
-\frac{(\Lambda c)^2}{2\pi^2}\intxy g_{3} (x,y) \cos\left[\sqrt{4
\pi}(\varphi(x) + \varphi (y)) -2i g (\Phi(x) + \Phi(y))\right].
\end{multline}

It is now convenient to  diagonalize the quadratic part of the
effective action by introducing the fields $\zeta$, $\chi$ and
$\phi$:

\begin{align}\label{eq:changeVariables}
\Phi&=\frac{i\zeta}{\tilde{g}}+\frac{2i\tilde{g}Bp^2}{\Delta+2B\tilde{g}^2p^2}\phi\\
\eta&=\frac{-iC}{2B\tilde{g}}\zeta -\frac{i\tilde{g}Cp^2}{\Delta +
2B\tilde{g}^2p^2}\phi +\frac{1}{\tilde{g}}\chi \\
\varphi&= - \zeta+\frac{\Delta}{\Delta+2B\tilde{g}^2p^2}\phi,
\end{align}
where we have defined $\tilde{g}^2=g^2/\pi$ and $\Delta(p)=C(p)^2-4A(p)B(p)$. We then
obtain

\begin{equation}
S_{0}=\frac{1}{2}\intp\left[\zeta(p)\left(p^2+\frac{\Delta}{2B\tilde{g}^2}\right)\zeta(-p)
+\chi(p)\frac{2B}{\tilde{g}^2}\chi(-p)+
\phi(p)\frac{p^2\Delta}{\Delta + 2B\tilde{g}^2p^2}\phi(-p)\right],
\label{44}
\end{equation}

\begin{multline}
S_{int}=-\frac{(\Lambda c)^2}{2\pi^2}\intxy g_{1} (x,y)
\cos\sqrt{4\pi}[\phi(x) -\phi (y)]\\-\frac{(\Lambda
c)^2}{2\pi^2}\intxy g_{3} (x,y) \cos\sqrt{4\pi}[\phi(x) + \phi
(y)].\label{45}
\end{multline}

One can see that the $\zeta$ and $\chi$ fields become completely
decoupled from $\phi$. Moreover, it becomes apparent that the
$\phi$-dependent piece of the action $S_{int}$ is the only one
containing potentially relevant contributions (i.e. gapped modes).

\section{RG treatment of the non-local umklapp coupling}
\setcounter{equation}{0} \hs In this Section we shall focus our
attention on the non local action derived above. For simplicity,
from now on we will consider the case in which $g_{2}(p)$ and
$g_{4}(p)$ are constants (local forward scattering) and $g_{1}=0,
g_{3}\neq 0$, i.e. a pure non local umklapp interaction. Thus, we
start from the action
\begin{multline}
S[\phi]=\intp\phi(p)\frac{F(p)}{2}\phi(-p) -\frac{(\Lambda
c)^2}{2\pi^2}\intxy
g_3(x,y)\cos\sqrt{4\pi}[\phi(x)+\phi(y)]\label{46}
\end{multline}
with

\begin{align}
F(p)=&\frac{1}{Kv}(p_0^2+v^2p_1^2)\\ K=&\sqrt{\frac{1+ g_{4}/ \pi
-g_{2}/ \pi}{1+ g_{4}/ \pi + g_{2}/ \pi}}\\
v=&\sqrt{\left(1+\frac{g_4}{\pi}+\frac{g_2}{\pi}\right)
\left(1+\frac{g_4}{\pi}-\frac{g_2}{\pi}\right)},\label{47}
\end{align}
where we have now expressed all formulae in terms of $g$ coupling
functions. In the local case ($g_{3}=\delta^{2}(x-y)$) the action
(\ref{46}) corresponds to the well-known sine-Gordon model, which
is an integrable, exactly solvable field theory. In particular, a
RG analysis shows that the ``stiffness constant", $K$ has to be
lower than $.5$ in order to have a relevant cosine interaction,
i.e. to have a gap in the spectrum. Recently, by reinterpreting
Bethe-Ansatz results, Zamolodchikov obtained the exact expression
for this gap \cite{Zamolodchikov}. Unfortunately, as far as we know, the present non
local version of the theory is not exactly soluble and one is then
forced to consider an approximation. In ref. \cite{Nosotros2} a
SCHA was employed in order to estimate the energy gap. But this is
a non controlled, non perturbative approximation. Besides this
general disadvantage, the implementation of the SCHA technique led
to a set of coupled algebraic equations that could be numerically
solved only for very weak non locality. It is then natural to try
another method to attack the problem and eventually improve the
approximation. Let us consider the Wilsonian approach to the RG
(See for instance \cite{Shankar}). First of all we restrict our
analysis to a non local interaction of the form:
\begin{equation} g_{3}(x - y) = g_{3}(x_{1} - y_{1})\delta(x_{0} -
y_{0}),\end{equation} with
\begin{equation}\label{g3}
g_3(x_{1})=\lambda_0\delta(x_{1})-\frac{\epsilon_{0}}{\Lambda^2}\partial_1^2\delta(x_{1})
\end{equation}
where $\partial_1^2\delta^2(x)$ is the second derivative of the
delta function with respect to $x_1$. At this point it is worth
mentioning that RG calculations involving fermionic non local
interactions already exist in the literature. For instance, the
authors of ref.\cite{SP} used a Jordan-Wigner fermion
representation for the 1D Heisenberg-Ising model which includes
not only a non local fermion-fermion interaction but also a linear
fermion-lattice coupling. Due to the curvature of the band, the
non trivial fermion-lattice coupling and the presence of both
forward and umklapp scattering, comparison of their RG equations
with ours is not straightforward. Let us point out, however, that
in \cite{SP} fermions corresponding to nearest-neighbor lattice
sites interact through a potential of the form $g(q)= C_{1} \times
\cos{q}$, i.e. a unique constant $C_{1}$ is associated to $g(q)$.
As shown in the above equation (\ref{g3}), in the present work we
are interested in a coupling which depends at least on two
constants $\lambda$ and $\epsilon$. In fact, the derivation of RG
equations for these coupling constants will be our next task.

In condensed matter problems one is usually interested in the
physics at long distances, compared to a lattice spacing of the
order of $\Lambda^{-1}$. Since, in momentum space, this
corresponds to small $k_1={\bf k}$, it is natural to consider
correlations between fields with momenta $0<k<\Lambda/s$, with s
very large. These are the so called "slow modes" $\phi_{<}$. On
the other hand, the "fast modes" $\phi_{>}$ are those carrying
momenta that satisfy $\Lambda/s \leq k \leq \Lambda$. In the
present approach to RG these fast modes are integrated in the path
integral framework, giving rise to an effective theory depending
only on slow modes. From this action, in turn, one can read the
flow equations for the couplings. Indeed, writing the initial
action as $S_0 - S_{int}$, to first order, after a suitable
rescaling of coordinates and momenta and a redefinition of the
fields (see the Appendix for details) we obtain the following
relationship between the original and RG transformed actions:
\begin{multline}\label{Sint} S_{int} = \frac{(\Lambda
c)^2}{2\pi^2}\int d^{2}x \,d^{2}y \left([\lambda_{0} +
\lambda_{0}\,( 2 - 4 K )t - 2 K \epsilon_{0} t
]\delta^2(x-y)-\frac{\epsilon - 4 K \epsilon
t}{\Lambda^2}\,\partial_1^2\delta^2(x-y) \right)\times \\ \times
\cos\sqrt{4\pi}(\phi(x) + \phi (y)) = \\ = \frac{(\Lambda
c)^2}{2\pi^2}\int d^{2}x\,d^{2}y\left(\lambda \delta^2(
x-y)-\frac{\epsilon}{\Lambda^2}\,\partial_1^2\delta^2(x-y)\right)
\cos\sqrt{4\pi}(\phi(x) + \phi (y)) = S_{int}^{'}.\end{multline}

As usual, imposing the invariance of the action under RG, we get
the flow equations for the couplings $\lambda$ and $\epsilon$:

\begin{equation}\label{rg1}\frac{d \lambda}{d t} = ( 2 - 4 K)\lambda - 2
K \epsilon \end{equation}

\begin{equation}\label{rg2} \frac{d \epsilon}{d t} = - 4 K \epsilon,
\end{equation}
where $\ln s = t$, with the initial conditions
$\lambda(0)=\lambda_{0}$ and $\epsilon(0)=\epsilon_{0}$. The
solution of this system is elementary, yielding:
\begin{equation}\label{rg3} \lambda(t) = (\lambda_{0} - K
\epsilon_{0})exp[(2 - 4 K)t] + K \epsilon_{0}\exp[-4 K
t],
\end{equation}
\begin{equation} \epsilon(t) = \epsilon_{0}\,exp[-4 K t].
\end{equation}
From the last equation one clearly sees that the non local piece
of the interaction is irrelevant, as expected. Concerning the
local interaction one sees that it is relevant for $K < 1/2$, i.e.
non locality does not modify this well-known condition already
found for the local SG model. Therefore, for $K < 1/2$, $\lambda$
will grow with increasing t and there will be a gap in the CD
spectrum which can be estimated by determining the value
$\tilde{t}$ for which $\lambda =1$. From now on we shall restrict
our analysis to the case $K < 1/2$. The gap is then given by
$\Delta E=\Lambda \mu \,e^{-\tilde{t}}$, where $\mu=c/(\pi
\sqrt{2})\approx .198$. It is also convenient to define the
dimensionless gap $m=\Delta E/\Lambda$. From equation (\ref{rg3})
one can thus derive an equation for $m$ which gives the behavior
of the energy gap as function of the forward scattering potentials
($K$) and the non local contribution of the umklapp scattering
($\epsilon_{0}$). Before analyzing this non trivial equation it
seems reasonable to check if it predicts sensible results for the
local case. To this end we set $\epsilon_{0}=0$ and $\lambda=1$ in
(\ref{rg3}), obtaining:
\begin{equation}
m_{0}=\mu\, \lambda_{0}^{\frac{1}{2-4K}}.
\end{equation}
This result can be compared with the exact solution obtained by
Zamolodchikov \cite{Zamolodchikov} and with the approximated
result given by the SCHA method \cite{Nosotros2}. The
corresponding expressions for the gap are respectively given by:
\begin{equation}
m_{Z}=\frac{2^{\frac{-2K}{1-2K}}}{\sqrt{\pi}}\,\mu^{\frac{1}{1-2K}}\,\frac{\Gamma(\frac{K}{1-2K})}{\Gamma(\frac{1}{2-4K})}\,
\left(\pi\,\frac{\Gamma(1-2K)}{\Gamma(2K)}\right)^{\frac{1}{2-4K}}
\,\lambda_{0}^{\frac{1}{2-4K}}
\end{equation}
and
\begin{equation}
m_{scha}=\frac{2}{\sqrt{K}}\,\left(4\pi\,K
\right)^{\frac{1}{2-4K}}\,\mu^{\frac{1}{1-2K}}\,
\lambda_{0}^{\frac{1}{2-4K}}.
\end{equation}
In order to compare $m_{0}$ and $m_{scha}$ with $m_{Z}$ in an
efficient and easy to visualize way, we have computed the relative
error $\Delta m/m_{Z}$ as function of $K$ for both approximations.
The results are depicted in Fig.1 where one sees that our RG
computation gives values of the gap closer to the exact values for
a wide range of the stiffness constant $K$. Interestingly, the
SCHA result works well when one approaches the end points of the
interval.

Going back to the case $\epsilon_{0} \neq 0$, by combining
eqs.(\ref{rg1}) and (\ref{rg2}) one readily gets a phase diagram
in the $\epsilon - \lambda$ plane (See Fig. 2). There is a
critical line given by $\lambda = K\,\epsilon$. If the initial
parameters are tuned to lie on this line, the system will flow to
the Tomonaga-Luttinger fixed point, at the origin. In this case,
of course, the system remains gapless. On the other hand, for
initial conditions outside the critical line, the system flows to
strong coupling, giving rise to a gap $m$, as mentioned above. For
simplicity let us consider the case $\lambda > \epsilon$ and
define the variable $x = m/m_{0}$. The gap equation can then be
written as
\begin{equation}\label{gap1}
x^2 - \frac{\nu}{\epsilon_{0}}\,x^{2-4K} +
\frac{\nu}{\epsilon_{0}}\,(1-\frac{K\epsilon_{0}}{\lambda_{0}})=0,
\end{equation}
where $\nu=\lambda_{0}^{\frac{-2K}{1-2K}}/K$. This equation is one
of our main results. For fixed values of $K$ and $\lambda_{0}$ it
gives the behaviour of the energy gap as function of the non local
contribution to umklapp scattering, associated to non contact
interactions. The form of this formula suggests that it could be
easier to handle the inverted equation:
\begin{equation}\label{gap2}
\epsilon_{0}(x)=\epsilon_{crit}\,\nu\,\frac{1-x^{2-4K}}{\nu-\epsilon_{crit}\,x^2},
\end{equation}
where $\epsilon_{crit}=\lambda_{0}/K$. In Fig.3 we show the
numerical solution of this last equation for $\lambda_{0}=.5$ and
$K=.25$. We see that $x$ decreases for increasing $\epsilon_{0}$,
in qualitative agreement with the SCHA prediction
\cite{Nosotros2}. A quantitative comparison between both
approximations is given in Fig.4, for the same fixed values of
$\lambda_{0}$ and $K$. Since the SCHA result obtained in
\cite{Nosotros2} is valid for small values of the coefficient
associated to non locality, we have plotted the solutions in the
interval $0 \leq \epsilon_{0} \leq 0.1$. We find that the gap
decay predicted by RG, for increasing non locality, is much slower
than the one obtained through the Gaussian approximation.

As a final comment, we note that the RG treatment for non local
interactions depicted in this work can be extended to a more
general coupling function including an arbitrary number of even
powers of $p_1$. In coordinate space such an interaction can be
written as
\begin{equation}
g_{3}(x) = \sum^{n}_{i = 0} (-1)^{i}\frac{\lambda_{i}}{\Lambda^{2
i}}\partial^{2i}\delta(x)\,
\end{equation}
where $\lambda_{1}=\epsilon_0$. The corresponding set of RG
equations for the $\lambda_{i}'s$ can be obtained from the
expansion of the cosine integral function which appears when one integrates the fast modes
after the mode separation (See Appendix). The result is

\begin{multline}\frac{d \lambda_{0}}{d t} = ( 2 - 4 K)\lambda_{0} - 2
K \sum^{n}_{i = 1}\lambda_{i} \\ ......... \\ \frac{d
\lambda_{j}}{d t} = ( 2 - 2 j - 4 K)\lambda_{j} - 2 K \sum^{n}_{i
= j + 1}\lambda_{i} \\ ......... \\\frac{d \lambda_{n}}{d t} = ( 2
- 2 n - 4 K)\lambda_{n}.\end{multline}

The general solution of this system can be expressed as a
combination of exponentials and the computation of the gap cannot
be done, in principle, in an analytical way. It is then
illustrative to consider a particular case in which this
calculation is simplified. Indeed, for small $t$, it can be proved
by induction that the solutions of this system of equations are of
the form:

\begin{equation} \lambda_{j}(t) = \lambda_{j}^{0}\exp[t( 2 - 2 j -
4 K - \frac{2 K}{\lambda_{j}^{0}}\sum^{n}_{i = j +
1}\lambda_{i}^{0})].\end{equation}
In this limit one obtains the following expression for the gap:
\begin{equation} m = \mu \lambda_{0}^{\frac{1}{2 -
4 K - \frac{2 K}{\lambda_{0}^{0}}\sum^{n}_{i =
1}\lambda_{i}^{0}}},\end{equation}
which is consistent with the conditions
$\lambda_{0}\sim 1^{-}$ and $K \ll \frac{1}{2}$.

\section{Conclusions}
\setcounter{equation}{0}

In this paper we have considered a non local extension of the
sine-Gordon model. This theory is obtained when one bosonizes a
non local and non covariant version of the Thirring model used to
describe certain 1d many-body systems. Since the integrability of
this non local sine-Gordon model has not been proved, one needs to
implement some approximation to study its physical content. We
have performed a RG calculation up to first order in the coupling
function $g_3$, which in a condensed matter context is associated
to the so called umklapp scattering. We obtained an expression for
the energy gap as function of the non local piece of the
interaction $\epsilon_0$. For purely local interactions (the
exactly solvable SG) our result seems to be a sensible
approximation, improving the SCHA predictions for a wide range of
forward interactions. In the non local case, in which no exact
answer is known, we predict decreasing values for the gap for
increasing values of $\epsilon_0$, in qualitative agreement with a
previous SCHA computation. We were able to give a precise
comparison between both approximations in the interval $0\leq
\epsilon_0 \leq 0.1$, showing that the gap decrease, for
increasing non locality, is much weaker according to the RG
computation. Since, as is well-known, the SCHA method is not a
controlled approximation, the present results contribute to a
better understanding of the physics of non local field theories.
We think that our results are also of interest in the context of
1D many-body systems (Luttinger liquids) in which most of the
previous investigations involving umklapp scattering do not
consider non local effects associated to long range interactions
\cite{um-news}.

\section*{Acknowledgements}

This work was partially supported by Universidad Nacional de La
Plata  and Consejo Nacional de Investigaciones Cient\'{\i}ficas y
T\'ecnicas, CONICET (Argentina).

\section{Appendix}

\hs In order to illustrate the computation leading to the flow
equations we first define the free bosonic propagator:

\begin{equation} <\phi(k)\,\phi(q)>_{0} = \frac{1}{Z_{0}} \int D
\phi \, \phi(k)\,  \phi(q)\, e^{- S_{0}} =
\frac{1}{Z_{0}}\frac{\delta^{2} Z(j)}{\delta j(k)\, \delta
j(q)}\mid_{j = 0}\end{equation} with
\begin{equation} Z(j) = \int D
\phi\,\exp[- \int\,\frac{d^{2}p}{2 (2\pi)^{2}}\,
(\Phi(p)\,F(p)\,\phi(-p) - J(p)\,\phi(p)) ].\end{equation} The
result is
\begin{equation} <\phi(p)\,\phi(q)>_{0} = \frac{\delta^{2}( p + q
)\,4 \pi^{2}\,K\, v }{p_{0}^{2} + v^{2}p_{1}^{2}}.\end{equation}

The next step is the analysis of $S_{int}$, as given by the second
term of equation (\ref{46}). For simplicity, in this Appendix we
disregard the overall constant $(\Lambda c)^2/(2\pi^2)$. Going to
momentum space and performing the separation in slow and fast
modes $\phi_{<}$ and $\phi_{>}$, according to:

\begin{equation} \phi(x) =
\frac{1}{(2\pi)^{2}}\int\,d^{2}p\,\phi(p)\,e^{ip.x}\end{equation}
\begin{equation} \phi(p) = \phi_{<}(p) , |p_{1}|< \Lambda/s
\end{equation}
\begin{equation} \phi(p) = \phi_{>}(p) , \Lambda/s \leq |p_{1}| \leq
\Lambda,
\end{equation}
we obtain
\begin{multline} S_{int} = \int\,d^{2}x\,dy_{1}\,g_{3}(x_{1} -
y_{1})(\cos[\frac{\sqrt{4\pi}}{(2\pi)^{2}}\int\,d^{2}p\,\phi_{<}(p)\,f(p,x,y_{1})]
\\ \cos[\frac{\sqrt{4\pi}}{(2\pi)^{2}}\int\,d^{2}p\,\phi_{>}(p)\,f(p,x,y_{1})] - \sin[\frac{\sqrt{4\pi}}{(2\pi)^{2}}\int\,d^{2}p\,\phi_{<}(p)\,f(p,x,y_{1})]
\\ \sin[\frac{\sqrt{4\pi}}{(2\pi)^{2}}\int\,d^{2}p\,\phi_{>}(p)\,f(p,x,y_{1})])\end{multline}
where
\begin{equation}f(p,x,y_{1}) = e^{ip.x} + e^{i(p_{0}x_{0} +
p_{1}x_{1})}.\end{equation}

Now we expand the functional integral up to first order in $g_3$
and integrate over the fast modes. One finds the following
results:

\begin{equation}<\sin[\frac{\sqrt{4\pi}}{(2\pi)^{2}}\int\,d^{2}p\,\phi_{>}(p)\,f(p,x,y_{1})]>_{0}
= 0
\end{equation}
and

\begin{multline}<\cos[\frac{\sqrt{4\pi}}{(2\pi)^{2}}\int\,d^{2}p\,\phi_{>}(p)\,f(p,x,y_{1})]>_{0}\,
=\\ =\exp \left(-K \int\,dp_1
\,(\theta(p_1)-\theta(-p_1))\,\frac{1+\cos(p_1(x_1-y_1))}{p_1}\right)=\\
=\exp \left(-K \left[2\ln s + 2 Ci\,(\Lambda(x_1-y_1))-2 Ci
(\frac{\Lambda}{s}(x_1-y_1))\right]\right)
\end{multline}
where $Ci(x)$ is the cosine integral function and the free vacuum
expectation values  are, of course, taken with respect to fast
modes. Rescaling momenta, coordinates and fields such that the
free piece of the action $S_0$ remains invariant:
\begin{equation} p^{'} = s p\, ,\, x^{'} = s^{-1} x\, ,\, \phi^{'}(p^{'})
= s^{-2}\,\phi_<(p^{'}/s),\end{equation} and using the fact that
$\ln s = t , s \simeq 1 + t$, $S_{int}$ can be written as
\begin{multline}S_{int} = \int\,d^{2}x^{'}\,dy_{1}^{'}\,g_{3}(s (x_{1}^{'}- y_{1}^{'}))\cos[\frac{\sqrt{4\pi}}{(2\pi)^{2}}\int\,d^{2}p\,\phi_{<}(p)\,f(p,x,y_{1})]\times
\\ \times\,( 1 + [ ( 3 - 4 K)  + K (\Lambda (x_{1}^{'}- y_{1}^{'}))^{2} -
\frac{K}{12}(\Lambda (x_{1}^{'}- y_{1}^{'}) )^{4}] t ),
\end{multline}
where we have used the power expansion of the function $Ci(x)$.
Finally, using the explicit expression for $g_3$ in terms of
$\lambda_0$ and $\epsilon_0$ we obtain

\begin{multline} S_{int} =\int d^{2}x^{'}\,d^{2}y^{'}[(\lambda_0 + \lambda_0 \,[ 2 - 4 K ]t - 2 K
\epsilon_0 t )\delta^2(x_{1}^{'}- y_{1}^{'})-\\-\frac{\epsilon_0 -
4 K \epsilon_0 t}{\Lambda^2}\partial_1^2\delta^2(x_{1}^{'}-
y_{1}^{'})] \cos\sqrt{4\pi}[\phi(x^{'}) + \phi (y^{'})] = \\ =
\int d^{2}x^{'}\,d^{2}y^{'}[\lambda^{'} \delta^2(x_{1}^{'}-
y_{1}^{'})-\frac{\epsilon^{'}}{\Lambda^2}\partial_1^2\delta^2(x_{1}^{'}-
y_{1}^{'})] \cos\sqrt{4\pi}[\phi(x^{'}) + \phi (y^{'})] =
S_{int}^{'},\end{multline} which leads to the flow equations for
$\lambda$ and $\epsilon$.

\newpage

{\bf Figure captions}\\

Figure 1: Relative error $\Delta m/m$ as function of the stiffness
constant, for the local case. The dashed line corresponds to the
SCHA whereas the solid line shows the RG result.
\\

Figure 2: Phase diagram in the $\epsilon-\lambda$ plane, for
$K=.25$. The dashed line is the critical line $\lambda =
K\,\epsilon$.
\\

Figure 3: Numerical solution of the gap equation for
$\lambda_{0}=.5$ and $K=.25$. The gap $m$ decreases for increasing
$\epsilon_{0}$.
\\

Figure 4: Numerical solutions of the gap equations for
$\lambda_{0}=.5$, $K=.25$, and $0 \leq \epsilon_{0} \leq 0.1$. The
dashed line corresponds to the SCHA whereas the solid line shows
the RG result. The unit function is included to allow comparison
with the local case.

\end{document}